\documentclass[aps,pre,floatfix,twocolumn,nofootinbib,showpacs]{revtex4}
\usepackage{graphicx} 
\usepackage[sort&compress]{natbib}
\usepackage{graphicx}
\newcommand{\BEQ}{\begin{equation}}
\newcommand{\EEQ}{\end{equation}}
\newcommand{\BEA}{\begin{eqnarray}}
\newcommand{\EEA}{\end{eqnarray}}

\renewcommand{\d}{{\rm d}}

\newcommand{\half}{\frac{1}{2}}

\newcommand{\Tr}{{\rm Tr}}

\newcommand{\CF}{{\cal F}}
\newcommand{\CZ}{{\cal Z}}
\newcommand{\CS}{{\cal S}}
\newcommand{\CU}{{\cal U}}
\newcommand{\Ts}{T_{\rm s}}

\newcommand{\Ss}{{\cal S}_{\rm s}}
\newcommand{\Stot}{{\cal S}_{\rm tot}}
\newcommand{\thetas}{\theta_{\rm s}}
\begin{document}
\draft
\title
{Anomalous latent heat in non-equilibrium phase transitions}

\author{
A.E. Allahverdyan and K.G. Petrosyan}
\address{Yerevan Physics Institute,
Alikhanian Brothers St. 2, Yerevan 375036, Armenia
}

\date{\today}

\begin{abstract} We study first-order phase transitions in a
two-temperature system, where due to the time-scale separation all the
basic thermodynamical quantities (free energy, entropy, etc) are
well-defined.  The sign of the latent heat is found to be
counterintuitive: it is positive when going from the phase where the
temperatures and the entropy are higher to the one where these
quantities are lower.  The effect exists only out of equilibrium and
requires conflicting interactions.  It is displayed on a
lattice gas model of ferromagnetically interacting spin-1/2 particles.

\end{abstract}

\pacs{64.60.Cn, 05.70.Ln, 75.10Hk}




\maketitle

The theory of equilibrium phase transitions is an established field with
known achievements in describing transformations of various states of
the matter \cite{landau,bell}. Concepts borrowed from this theory apply
for some non-equilibrium transitions which get equilibrium features on
the macroscopic scale, e.g., equilibrium universality classes for
non-equilibrium transitions \cite{zia}, transitions to the glassy state
analyzed with help of effective temperature \cite{cug,th}, etc. Less is
known, however, about phase transition scenarios which are impossible in
equilibrium.

Here we study such a truly non-equilibrium phase transition scenario
realized in the steady state of a two-temperature system.  This
system admits a natural generalization of the equilibrium
statistical thermodynamics, i.e., the quantities like entropy,
internal energy, free energy are well defined, because the
constituents of the system are in local equilibrium.  In spite of
that, the system shows a counterintuitive type of first-order phase
transition, where the latent heat is positive ({\it anomalous latent
heat}) when transforming a high temperatures (higher entropy) phase
to the low temperature (lower entropy) one, i.e., in the first phase
the energy is larger.  This is a non-equilibrium effect. It is
well-known, even from the everyday physics, that for equilibrium
transitions the latent heat is negative \cite{landau}.  This plays a
crucial role in the heat balance of the Earth and in the weather
formation, since nearly 70\% of the energy transferred from the Earth's
surface is due to the latent heat consumed during the vaporization
at the surface and released during the vapor condensation in the
atmosphere \cite{climat}.

First we shall recall the thermodynamics of two-temperature systems with
different time-scales \cite{LandauerWoo,Onsager,Ritter}. 
In contrast to the usual equilibrium case, the anomalous
latent heat is not forbidden here. We then work out a simple model of
mean-field Ising ferromagnet demonstrating the sought effect.  A
necessary condition for its existence is the presence of conflicting
interactions.

Consider a pair of coupled stochastic variables $s$ and $f$ with
Hamiltonian $H(s,f)$, which interact with different thermal baths at
temperatures $\Ts$ and $T_f\equiv T$, respectively ($s$ and $f$ can
denote a set of variables).  For $T=\Ts=1/\beta$, the stationary
probability distribution $P(s,f)$ of the system is Gibbsian:
$P(s,f)\propto e^{-\beta H(s,f)}$. For
$T\not=\Ts$ we can derive the
stationary $P(s,f)$ if the variables have different characteristic
times: $s$ is slow, while $f$ is fast (adiabatic limit). This derivation
together with corrections coming from a large, but
finite time-scale difference was given in \cite{Onsager} based on
stochastic equations of motion. Here we recall the heuristics
\cite{LandauerWoo,Onsager,Ritter,Coolen}.  On the times relevant for $f$, $s$ is fixed,
and the conditional probability $P(f|s)$ is Gibbsian:
\BEA
\label{bora77}
P(f|s)=\frac{1}{Z(s) }{e^{-\beta H(s,f)}},
~~ Z(s) = \Tr_f \,e^{-\beta H(s,f)  }
\EEA
where $\Tr_f$ is the sum over all values of $f$.  The steady-state
$P(s)$ is found by noting that on the times relevant for $s$, $\tau $ is
already in the conditional steady state. Thus the force $ \partial_s
H(s,f)$ acting on $s$ can be averaged over $P(f|s)$: $\Tr_f \left[ \partial_s
H(s,f) P(f|s)\right ]=\partial_s F(s) $, where $F(s)=-T\ln Z(s)$
is the conditional free energy. The steady $P(s)$ is Gibbsian with the Hamiltonian
$F(s)$:
\BEA
\label{bora88}
P(s)=\frac{e^{-\beta_{\rm s} F(s)}}{\CZ  }=\frac{Z^{\frac{T}{\Ts}} (s)   }{\CZ  },\quad
\CZ=\Tr_s  e^{-\beta_{\rm s} \CF(s)} ,
\EEA
and the common probability is $P(s,f)=P(s)P(f|s)$.

This two-temperature situation admits a (generalized)
thermodynamical description, because in the adiabatic limit both
variables are in local equilibrium \cite{LandauerWoo,Onsager,Ritter}.  (It has certain
analogies with non-equilibrium thermodynamics of the glassy state
proposed in \cite{th}.) The average energy of the system is
$\CU=\Tr_{s,f} \left[P(s,f)H(s,f)\right]$, or \BEA \label{pamir}
\CU=-\frac{1}{n}\partial_\beta \ln \CZ|_{n},~~n\equiv\frac{T}{\Ts}
\EEA where the derivative is taken for fixed $n$ and we used
Eqs.(\ref{bora77}, \ref{bora88}). $\CZ$ is a generating function,
since an arbitrary average is found via the proper
source term analogously to (\ref{bora88}). The entropies of $s$ and $f$ are, respectively, \BEA
\label{kaban1}
&&\Ss=-\Tr_s P(s)\ln P(s),\\
\label{kaban2}
&&\CS=-\Tr_s P(s)\left[
\Tr_f P(f|s)
\ln P(s|f)
\right].
\EEA
Eq.(\ref{kaban1}) is the usual definition of entropy. Eq.(\ref{kaban2})
is the conditional entropy; it appears due to the adiabatic limit.
The total entropy is the sum of partial ones
$\Stot\equiv -\Tr_{s,f} \left[P(s,f)\ln P(s,f)\right]=\CS+\Ss$.
Now the steady distributions
(\ref{bora77}, \ref{bora88}) are obtained when minimizing
the free energy of the slow motion
\BEA
\label{freedom}
\CF=-\Ts\ln \CZ.
\EEA
To this end, note from (\ref{bora77}--\ref{kaban2})
that the free energy and the entropies are expressed as
\BEA
\label{chin}
&&\CF=\CU-T\CS-\Ts\Ss,\\
\label{min}
&&\Ss=-\partial_{\Ts} \CF|_{T},
~~\CS=-\partial_{T} \CF|_{\Ts},
\EEA
generalizing the usual thermodynamical relations.

Let us now study the first-order phase transitions in this
approach.  Consider two phases $l$ and $h$.  Assume that when
$\Ts$ decreases below some critical value $\Ts^{(\rm c)}$
($T$ is fixed), the phase $l$ dominates, since its free energy
is smaller: $\CF_l<\CF_h$.  For $\Ts>\Ts^{(\rm c)}$ the dominating
phase is $h$: $\CF_l>\CF_h$.  At high (low) temperatures $\Ts$, $l$
($h$) is metastable. $\CF_l=\CF_h$ at $\Ts=\Ts^{(\rm c)}$,
while other quantities change by jump.  Denoting $\delta X=X_l-X_h$ we get
\BEA
\label{tibet}
\delta \CU=\Ts\delta\Ss+T\delta \CS
=-\Ts\partial_{\Ts}\,[\delta \CF]|_T
-T\partial_T\,[\delta \CF]|_{\Ts},
\EEA
where $\delta\CU$ is the latent heat of the transition.
Note that in general both $\delta\Ss$ and $\delta\CS$ are
functions of $\Ts$ and $T$. In the vicinity
of $\Ts^{(\rm c)}$, $\delta\CF=\CF_l-\CF_h$ is an increasing function of $\Ts$
(for fixed $T$); thus $\delta\Ss<0$. The sign of $\delta\CS$ is,
however, left open and we cannot conclude that $\delta\CU<0$. It may be even
positive (anomalous latent heat), provided $\delta\CS>0$, that is
provided the system moves from $l$ to $h$ when decreasing $T$ for
constant $\Ts$.

Things get different if we decrease both $\Ts$ and $T$, with constant
$n=T/\Ts$. Since $\delta \CF(\Ts,n\Ts)=\CF_l-\CF_h$ has to be an
increasing function of $\Ts$, we get $\frac{\d}{\d \Ts}\delta\CF
=(\partial_{\Ts}+n\partial_T)\delta \CF>0$, and then (\ref{tibet}) shows
that $\delta\CU<0$ confirming the equilibrium result for $\Ts=T$
\cite{landau}.

Thus for $\Ts\not=T$ the anomalous latent heat is not forbidden,
provided the transition is not driven by a proportional change of
both temperatures.  Here is a model demonstrating this effect.
Particles are located at the nodes of a lattice embedded into a
particle reservoir with chemical potential $\mu=-\alpha<0$. Each particle
carries an Ising spin. At short distances the particles repel each
other so that not more than one particle can occupy a singe node.
The Ising spins interact ferromagnetically. This interaction is
active between two nodes only if they both are occupied;
$s\equiv\{s_i=\pm 1\}_{i=1}^N$ and $f\equiv\{f_i=0,1\}_{i=1}^N$ are,
respectively, the spins and the occupations of $N$ nodes. We assume
all nodes can interact with each other (mean-field). The Hamiltonian
reads \BEA H(s,f)= -\frac{J}{2N}{\sum}^N_{i\not=k} f_i f_k s_i s_k
+\alpha{\sum}_{i} f_{i}, \label{granada73} \EEA where $J/(2N)>0$ is
the ferromagnetic coupling constant with $J={\cal O}(1)$, as required
for the extensivity.

We make two assumptions. 1) $s$ and $f$ couple, respectively, to a spin
bath and lattice bath at different temperatures, $\Ts=\frac{1}{\beta_{\rm s}}$
and $T=\frac{1}{\beta}$. 2) $s$ ($f$) is slow (fast), i.e., the relaxation time
$\tau_f$ of $f$ is much shorter than the relaxation time $\tau_s$ of
$s$. These times are driven by the interactions with thermal baths, since
$H$ contains only commuting terms; note that in the quantum setting
$s_i=s_{i,z}$ and $f_i=\half(1+\sigma_{i,z})$, where $s_{i,z}$ and
$\sigma_{i,z}$ are the third Pauli matrices.

The above assumptions are motivated by NMR/ESR physics, where for
nuclear or electronic spins the spin bath is realized by relatively
weak dipole interactions and does indeed lead to a temperature
different from the lattice one \cite{goldman}. This spin temperature
can be tuned by external fields or by spin cooling and plays an
important role in analyzing experiments \cite{goldman}.  The
relaxation time $\tau_s$ on which the spin temperature is
established is known as $T_2$ time in NMR/ESR, and usually varies
between $10^{-4}$s and $1$s depending on the material.  The spin and
the lattice temperatures tend to equalize on the $T_1$ time, which
for many magnetic materials amounts to minutes or hours
\cite{goldman}. Estimating the relaxation time $\tau_f$ of $f$ via
the rotation-vibration mechanism as $10^{-12}-10^{-9}$s
\cite{goldman}, we see that our assumptions are based on
$\tau_f\ll\tau_s\ll T_1$.

The model is a suitable candidate for displaying the anomalous latent
heat, since the two terms in the RHS of (\ref{granada73}) are in
conflict: for two lined up spins, $s_is_k>0$, the term
$-\gamma{\sum}_{<i,k>} f_i f_k s_i s_k$ tends to increase occupations,
$f_if_k>0$, while $\alpha\sum_if_i$ makes this increase costly. The
equilibrium ($\Ts=T$) version of the model is related to Blume-Capel
model \cite{bell}. Similar models were employed recently for describing
solid mixtures \cite{bell}, inverse freezing (reentrance) phenomena
\cite{schupper} and glassy physics \cite{glassy}. 

Employing $e^{\left( \sum_if_is_i\right)^2}=c \int \d m\,
e^{-\frac{N}{2}\beta Jm^2+\beta J m\sum_is_if_i}$, where
$c=\sqrt{\frac{N\beta J}{2\pi}}$, we express from Eqs.(\ref{bora88}, \ref{granada73}) the
partition function $\CZ$ of the model as
\BEA
{\cal Z}=
{\rm Tr}_s\left[
{\rm Tr}_f \,e^{\frac{\beta J}{2N}
\left(\sum_if_is_i\right)^2-\alpha\beta\sum_if_i
}
\right]^n
=\int{\cal D}m
e^{-N\beta_s \CF},\nonumber
\label{zoro}
\EEA
where ${\cal D}m\equiv c^{\frac{n}{2}}\prod_{a=1}^n\int\d m_a$,
and we assumed that $n\equiv\frac{T}{T_s}$ is an integer (replica method); later
on we shall make continuation to real $n$.
We also defined $\phi(m)=1+e^{-\alpha\beta+\beta Jm}$ and
\BEA
\label{mani}
\beta_s \CF(m)=\frac{\beta J}{2}\sum_{a=1}^nm_a^2
-\ln\left[\sum_{k=\pm 1}
e^{\sum_{a=1}^n\ln \phi(km_a)}
\right].
\EEA
$\CZ$ is calculated by the saddle-point method, where
one searches the deepest minimum of $\CF$ as a function of $m_a$. This
minimum is reached for $m_a=m$ (replica symmetry), where $m$ satisfies the stationarity
condition $\frac{\partial }{\partial m_a} \CF|_{m_a=m}=0$, i.e.,
\BEA
\label{k2}
m=e^{-\alpha\beta}\,\frac{\phi^{n-1}(m)e^{\beta Jm}-\phi^{n-1}(-m)e^{-\beta Jm}}
{\phi^{n}(m)+\phi^{n}(-m)}.
\EEA
The corresponding value of $\CF$ obtained from (\ref{mani}) is
\BEA
\label{kaplan}
\CF=\frac{Jm^2}{2}-\Ts\ln\left[
\phi^{n}(m)+\phi^{n}(-m)
\right].
\EEA
It is seen that $\CF$ is the free energy of the system and
that $m$ is the magnetization:
\BEA
\label{bala}
m=\frac{1}{N}{\sum}_{i=1}^N\langle \tau_i\sigma_i\rangle.
\EEA
Due to the symmetry of the model, Eq.(\ref{k2}) admits two equivalent solutions with $\pm m$.
The choice between them is done spontaneously.

The energy $\CU$ and the entropies of the
spins $\CS_s$ and the occupations $\CS$ are calculated from
Eqs.(\ref{pamir}-\ref{granada73})
\BEA
\label{gomes1}
U=-\frac{Jm^2}{2}+\frac{\alpha}{z(m)} e^{-\alpha\beta}
{\sum}_{k=\pm 1}  \phi^{n-1}(km) e^{\beta Jkm},\\
\label{gomes2}
\Ss=\ln z(m)-\frac{n}{z(m)}
{\sum}_{k=\pm 1}\phi^n(km)\ln\phi(km),~~~~~\\
\label{gomes3}
\CS=\beta U+\frac{1}{z(m)} {\sum}_{k=\pm 1}
\phi^n(km)\ln\phi(km)-\frac{\beta Jm^2}{2},
\EEA
where $z(m)={\sum}_{k=\pm 1}\phi^n(km)$.
The two terms in RHS of (\ref{gomes1}) are equal, respectively, to
the spin-spin interaction energy
$\frac{J}{2N}\langle(\sum_if_is_i)^2\rangle$ and $\alpha {\cal N}$, where
${\cal N}=\sum_i\langle f_i\rangle$ is the average number of particles.

In the derivation of (\ref{gomes1}--\ref{gomes3}) we assumed
that the system is in the ferromagnetic phase:
$m\not=0$. For the paramagnetic phase $m=0$ one has from (\ref{kaplan})
\BEA
U=\frac{\alpha}{e^{\alpha \beta}+1},~~
\Ss=\ln 2,~~
\CS=\frac{\alpha\beta}{e^{\alpha \beta}+1}+\ln (1+e^{-\alpha \beta}).\nonumber
\EEA
The phase diagram of the model is constructed in terms of three adimensional parameters:
\BEA
a\equiv{\alpha}/{J},\quad \thetas={T_s}/{J},\quad \theta={T}/{J}.
\EEA
Let us start with
few particular cases. For $\theta\to 0$ we get from (\ref{kaplan})
(assuming $m\geq 0$)
\BEA
\label{mo}
\beta_{\rm s}\CF= \frac{m^2}{2\theta_s}-\ln[
1+\zeta(m-a)+\zeta(a-m)e^{(m-a)/\theta_s}
],\\
m=(1+e^{(a-m)/\theta_s})^{-1}(1-\zeta(a-m)),~~
\label{jo}
\EEA
where $\zeta(x)$ is approximated by the step function: $\zeta(x)=1$$(0)$
for $x>0$($x<0$). We see that $m\not=0$ only for $a<1$.  However, only
for $a<0.5$, Eq.~(\ref{jo}) predicts first-order phase transition to the
ferromagnet at $\theta_s=\theta_{\rm s}^{(c)}$.  Above this temperature
there are no particles in the system, ${\cal N}=0$, since the energy
cost for consuming a particle from the particle reservoir is too high
(due to $T\to 0$). At $\theta_s=\theta^{(c)}_{\rm s}$ a finite
fraction of particles is consumed from the particle reservoir making up
the ferromagnetic phase, which appears as a metastable state at a
higher critical temperature $\theta_{\rm s}^{*}>\theta_{\rm
s}^{(c)}$.  There is no first-order transition for $a>0.5$, though for
$1>a>0.5$ the ferromagnet can exist as a metastable phase.

\begin{figure}
\includegraphics[width=0.75\linewidth]{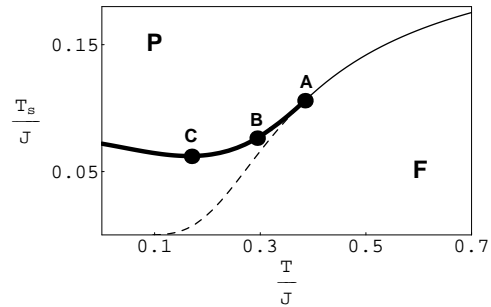}
\hfill
\caption{ The phase diagram on the
$T-\Ts$ plane; $\frac{\alpha}{J}=0.45$. {\bf P} and {\bf
F} refer to the paramagnet and ferromagnet, respectively.  Bold line:
first-order phase transitions. Normal line: second-order phase
transitions.  For $T\to\infty$ this line monotonically saturates at
$\Ts=0.25 J$, as seen from (\ref{saintgermain}).
Dashed line: the instability line of the paramagnet. On the AC segment the latent heat
is anomalous.
} \label{fig1}
\end{figure}
In a different limit $\alpha=0$ we get from (\ref{kaplan}):
\BEA
\label{otello}
\frac{\CF}{J}=\frac{m^2}{2}-\theta_s\ln\cosh(\frac{m}{2\theta_s})
-\theta\ln\cosh(\frac{m}{2\theta}),\\
\label{yago}
2m=\tanh(\frac{m}{2\theta})+\tanh(\frac{m}{2\theta_s}),
\EEA
Now there are only second-order phase transitions for the same reason as
for the usual (Curie-Weiss) mean-field ferromagnet: $\partial^3_mF$ can
never turn to zero for $m\not =0$. The phase diagram on the $\theta_{\rm
s}-\theta$ plane amounts to the ferromagnet (paramagnet) located below
(above) the line
\BEA
\theta^{-1}+\theta_{\rm s}^{-1}=4.
\label{saintgermain}
\EEA
The same Eqs.(\ref{otello}, \ref{yago}) without the terms containing
$\theta$ apply for $\theta\to \infty$. Here $f_i$ fluctuate so strongly
that the influence of $\alpha$ disappears.

A necessary condition for second-order transition is the local instability
of the paramagnet: $\partial^2_m\CF|_{m=0}=0$, i.e,
\BEA
\label{keta}
\theta_{\rm s}=[
1+e^{a/\theta}(2+e^{a/\theta}-\theta^{-1}
)]^{-1}.
\EEA
We could continue this reasoning and develop the Ginzburg-Landau
expansion, but here it is easier to study
(\ref{k2}, \ref{kaplan}) directly.
For $a>0.5$ there is
only the second-order phase transition to ferromagnet at temperatures
given by (\ref{keta}). Eq.(\ref{bala}) shows that
$m$ can increase by lining up the spins with or without increasing
the occupations.  For $a>0.5$ the second way works, since it is too
costly to absorb particles from the reservoir. Indeed, slightly below
the transition the difference in the average number of particles is small:
$\Delta{\cal N}\propto m^2\ll m$; see (\ref{gomes1}).

For $0.38<a<0.5$ the phase diagram is of the type presented in
Fig.\ref{fig1}.  Now there are first-order phase transitions from
the paramagnet to the ferromagnet. Recall Eqs.(\ref{mo}, \ref{jo})
for understanding their mechanism. The transition (phase
coexistence) line is found by solving (\ref{bala}), looking for the
deepest minima of $\CF(m)$, and requiring the continuous change of
the free energy: $\CF(m)=\CF(0)$.  The ferromagnet corresponds to
$\CF(m)<\CF(0)$ (recall the saddle-point method.) Other quantities
of interest change by jump. The ferromagnet first appears as a
metastable phase above the first-order transition line, while the
paramagnet survives as metastable till the dashed line on
Fig.\ref{fig1}. Thus, in the vicinity of the first-order transition
line, both phases are locally stable. This bistability implies
hysteresis and memory: when changing the temperature the final state
of the system (para or ferromagnet) depends on its initial state.
Since both $m$ and ${\cal N}$ increase in the ferromagnet, $m$ jumps
to a non-zero value not simply due to lining up the existing spins,
but also due to absorbtion of additional particles from the
reservoir.

On the AC segment the system moves from the paramagnet to the
ferromagnet upon {\it increasing} $T$ or {\it decreasing} $\Ts$. Thus,
$\Ss$ is smaller in the ferromagnet, while $\CS$ is larger there:
$\delta\Ss<0$ and $\delta\CS>0$; see (\ref{tibet}).  This is necessary
for the existence of the anomalous latent heat effect; see (\ref{tibet})
and below. On the AB segment the latent heat is {\it anomalous}: when
decreasing $\Ts$ for a fixed $T$ the system moves towards the
ferromagnet and its internal energy {\it increases}. This is due to a
positive energy brought about by the particles coming from the
reservoir; see (\ref{gomes1}).  The maximal magnitude of the effect
reached in the middle of AC segment is
$\delta\CU\sim J/20$. Recall that the transition to the ferromagnet is
induced by the tendency of the spin interaction energy to decrease.
The latent $\delta\CU$ is a consequence of this tendency that for
$\delta\CU>0$ overcomes its cause.  On the whole first-order
transition line the total entropy $\Stot=\CS+\Ss$, given
by (\ref{gomes2}, \ref{gomes3}), jumps by a negative
amount: expectedly the ferromagnet is more ordered than the paramagnet.

At the point B the latent heat is zero, while the BC segment shows an
anomalous latent heat in a different scenario: upon {\it decreasing} $T$
the system goes to the paramagnet, where the energy is {\it higher}.
The main difference between the two scenarios is that now the low-$T$
phase (paramagnet) has higher total entropy $\Stot$: a first-order
transition to a higher entropy phase is induced by decreasing the
temperature $T$. This is impossible in equilibrium; see Eq.(\ref{tibet}).
\begin{figure}
\includegraphics[width=0.75\linewidth]{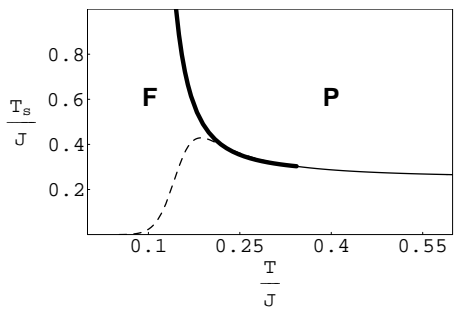}
\hfill
\caption{ The same as in Fig.(\ref{fig1}), but now ${\alpha}=0.245\,J$.
} \label{fig2} \end{figure}

An example of the phase diagram for $a<0.38$ is presented in
Fig.\ref{fig2}. The qualitative shape of the ferromagnet-paramagnet
boundary is given by (\ref{saintgermain}).  The transitions to the
ferromagnet are always induced when decreasing either temperature: there
is no anomalous latent heat here. During the first-order transitions the
particle number still increases: $\delta{\cal N}>0$, but now this brings to
decreasing entropy $\delta\CS<0$. In the present low-$a$ regime the occupations
are slaved by the spins so that the incoming particles arrange their
entropy in the way dictated by the spin-spin interaction energy.  For
$\frac{\alpha}{J}\to 0$, the first-order transitions gradually disappear.

In conclusion, we found first-order phase-transitions with an
unexpected sign of the latent heat: it is positive when going from
high temperature (high entropy) phase to the low temperature (low
entropy) one. Two conditions are necessary for this: global
non-equilibrium (two different temperatures), though the local
equilibrium is kept and leads to a slightly generalized
thermodynamics, and conflicting interactions.  The effect was
displayed on the model of lattice gas whose particles carry
ferromagnetically interacting Ising spins. Motivating by NMR/ESR
physics we assumed that the spins equilibrate at a temperature
different from the lattice one.  For this model there is a conflict
between attractive interaction facilitating the income of particles
from the reservoir and the chemical potential that makes this income
costly.

The effect seems to be generic, since we found it in many other situations,
e.g., in the present model, but with fast spins and slow
occupations, models with multi-spin interaction, the Ising spin-link ferromagnet, etc.

Thanks to Theo Nieuwehuizen for discussions.

A.E. A. was supported by CRDF grant ARP2-2647-YE-05.
K.G. P. was supported by ISTC grant A-820.

\end{document}